\documentclass[sigconf]{acmart}
\usepackage{todonotes}
\usepackage{enumitem}
\usepackage{graphicx}
\usepackage{epstopdf}
\usepackage{amssymb}
\usepackage{amsmath,amsthm}
\usepackage{comment}
\usepackage{longtable}
\usepackage{multirow}
\usepackage{booktabs}
\usepackage{colortbl}
\usepackage{tabularx}
\usepackage{lineno}
\usepackage{setspace}
\theoremstyle{definition}

\usepackage[font={small}]{caption}

\def\ie{{i.e.},~}
\def\eg{{e.g.},~}

\newcommand{\Curie}{{\textsc{\small{Curie}}}\xspace}
\iftrue
\newcommand{\berkay}[1]{{\color{red}{\bf BC:} #1}}
\else
\newcommand{\berkay}[1]{}
\fi

\usepackage{microtype}
\everypar{\looseness=-1}

 \usepackage[ruled,vlined,linesnumbered]{algorithm2e}
 \usepackage[ruled,vlined]{algorithm2e}
\usepackage{amssymb,amsmath}
\usepackage[hang,flushmargin]{footmisc}
\usepackage{lipsum}
\usepackage{syntax} 

 \makeatletter
\newcommand{\algorithmfootnote}[2][\footnotesize]{%
  \let\old@algocf@finish\@algocf@finish
  \def\@algocf@finish{\old@algocf@finish
    \leavevmode\rlap{\begin{minipage}{\linewidth}
    #1#2
    \end{minipage}}
  }
}
\makeatother

\usepackage{tikz}
\usepackage{pifont}
\newcommand{\cmark}{\ding{51}}%
\newcommand\inv[1]{#1\raisebox{1.15ex}{$\scriptscriptstyle-\!1$}}

\DeclareRobustCommand*\circled[1]{\tikz[baseline=(char.base)]{ \node[shape=circle,draw,color=white,fill=black,inner sep=0.5pt] (char){#1};}}
\newcolumntype{P}[1]{>{\centering\arraybackslash}p{#1}}

\newcommand{\xmark}{\ding{55}}

\usepackage[flushleft]{threeparttable}
\newcommand\numberthis{\addtocounter{equation}{1}\tag{\theequation}}
\usetikzlibrary{arrows,shadows,matrix,shapes,arrows,positioning,chains,calc}
\usepackage{pgf-umlsd}

\copyrightyear{2019} 
\acmYear{2019} 
\setcopyright{acmcopyright}
\acmConference[CODASPY '19]{Ninth ACM Conference on Data and Application Security and Privacy}{March 25--27, 2019}{Richardson, TX, USA}
\acmBooktitle{Ninth ACM Conference on Data and Application Security and Privacy (CODASPY '19), March 25--27, 2019, Richardson, TX, USA}
\acmPrice{15.00}
\acmDOI{10.1145/3292006.3300042}
\acmISBN{978-1-4503-6099-9/19/03}

\title{{\textsc{Curie}}: Policy-based Secure Data Exchange}
\author{Z. Berkay Celik}
\affiliation{SIIS Laboratory, Department of CSE\\The Pennsylvania State University}
\email{zbc102@cse.psu.edu}

\author{Abbas Acar, Hidayet Aksu}
\affiliation{CPS Security Lab, Department of ECE\\Florida International University}
\email{aacar001, haksu@fiu.edu}

\author{Ryan Sheatsley}
\affiliation{SIIS Laboratory, Department of CSE\\The Pennsylvania State University}
\email{rms5643@cse.psu.edu}

\author{Patrick McDaniel}
\affiliation{SIIS Laboratory, Department of CSE\\The Pennsylvania State University}
\email{mcdaniel@cse.psu.edu}

\author{A. Selcuk Uluagac}
\affiliation{CPS Security Lab, Department of ECE\\ Florida International University}
\email{suluagac@fiu.edu}

\begin{document}
\renewcommand{\shortauthors}{Celik et al.}
\begin{abstract}
Data sharing among partners---users, companies, organizations---is crucial for the advancement of collaborative machine learning in many domains such as healthcare, finance, and security.  Sharing through secure computation and other means allow these partners to perform privacy-preserving computations on their private data in controlled ways.  However, in reality, there exist complex relationships among members (partners). Politics, regulations, interest, trust, data demands and needs prevent members from sharing their complete data. Thus, there is a need for a mechanism to meet these conflicting relationships on data sharing. This paper presents \Curie\footnote{Our paper named after Marie Curie. She is physicist and chemist who conducted pioneering research in health care and won Nobel prize twice.}, an approach to exchange data among members who have complex relationships. A novel policy language, CPL, that allows members to define the specifications of data exchange requirements is introduced. With CPL, members can easily assert who and what to exchange through their local policies and negotiate a global sharing agreement. The agreement is implemented in a distributed privacy-preserving model that guarantees sharing among members will comply with the policy as negotiated. The use of \Curie is validated through an example healthcare application built on recently introduced secure multi-party computation and differential privacy frameworks, and policy and performance trade-offs are explored.
\end{abstract}



\settopmatter{printfolios=true}
\settopmatter{printacmref=false}

\maketitle

\section{Introduction}
Inter-organizational data sharing is crucial to the advancement of many domains including security, health care, and finance. Previous works have shown the benefit of data sharing within distributed, collaborative, and federated learning~\cite{dean2012large, smith2017federated, anil2018large}. 
Privacy-preserving machine learning offers data sharing among multiple members while avoiding the risks of disclosing the sensitive data (\eg health-care records, personally identifiable information)~\cite{el2013secure}. 
For example, secure multiparty computation enables multiple members, each with its training dataset, to collaboratively learn a shared predictive model without revealing their datasets~\cite{mohassel2017secureml}.
These approaches solve the privacy concerns of members during model computation, yet do not consider the complex relationships such as regulations, competitive advantage, data sovereignty, and jurisdiction among members on private data sharing.
Members want to be able to articulate and enforce their conflicting requirements on data sharing.

\begin{figure}[t!]
\begin{center}
\includegraphics[width=0.7\columnwidth]{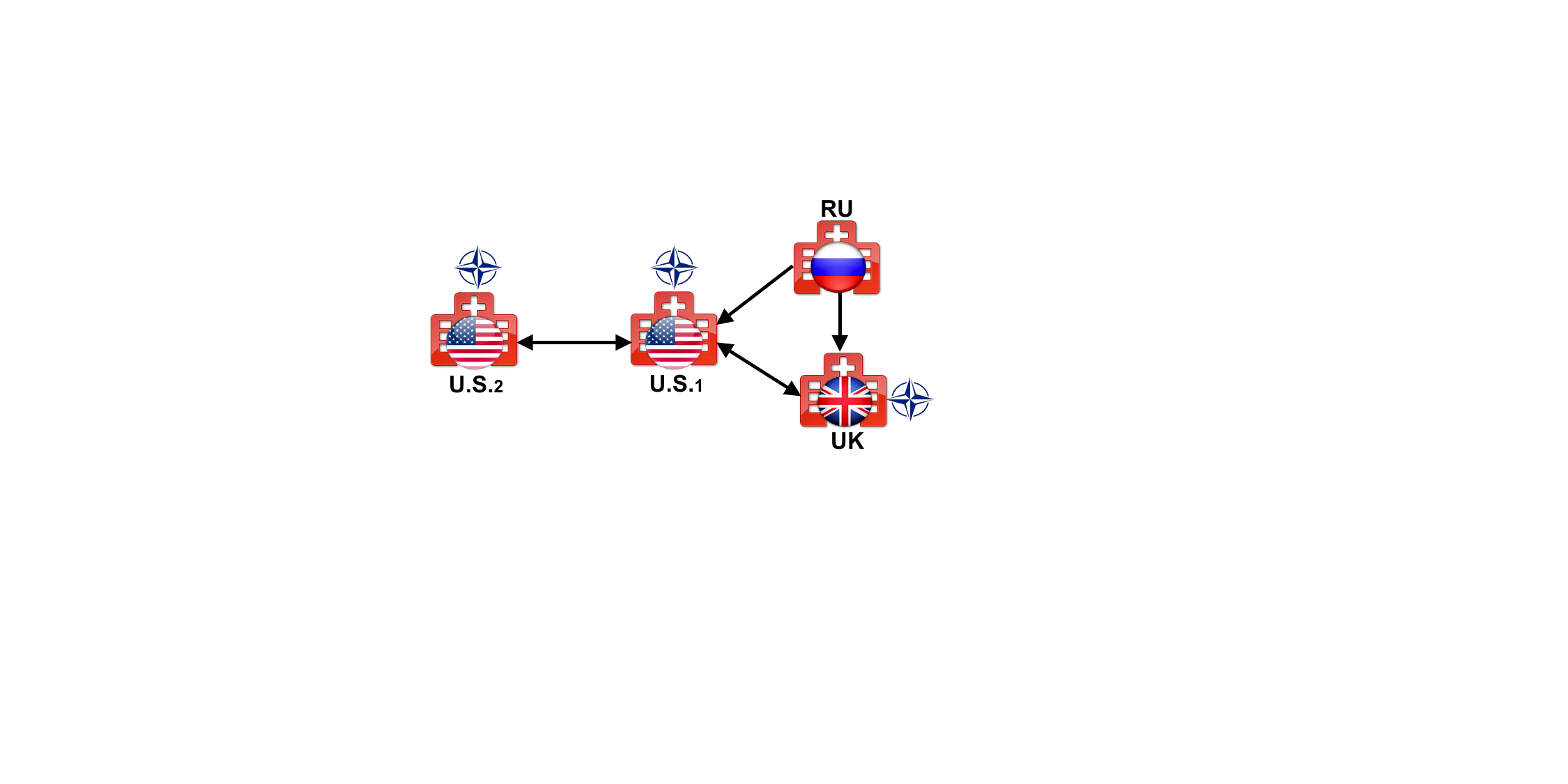}
\caption{An illustration of data exchange requirements of countries learning a predictive model on their shared data. Arrows show the data requirements of countries.}
\label{fig:motivation}
\end{center}
\end{figure}

To illustrate such complex data sharing requirements, consider health care organizations that collaborate for a joint prediction model of diagnosis of patients experiencing blood clots (see Figure~\ref{fig:motivation}). Members wish to dictate their needs through their legal and political limitations as follows: $\textrm{U.S.}_1$ is able to share its complete data for nation-wide members ($\textrm{U.S.}_2$)~\cite{arra, hitech}, yet it is obliged to share the data of patients deployed in NATO countries with NATO members ($\textrm{UK}$)~\cite{nato-share}. However, $\textrm{U.S.}_1$ wishes to acquire all patient data from other countries. $\textrm{UK}$ is able to share and acquire complete data from NATO members, yet it desires to acquire only data of certain race groups from $\textrm{U.S}_1$ to increase its data diversity. $\textrm{RU}$ wishes to share and acquire complete data from all members, yet members limit their data share to Russian citizens who live in their countries. Such complex data sharing requirements also commonly occur today in non-healthcare systems~\cite{lindell2009secure, solove2003information}. For instance, National Security Agency has varying restrictions on how human intelligence is shared with other countries; financial companies share data based on trust, and competition among each other.

This paper presents a policy-based data exchange approach, called \Curie, that allows secure data exchange among members that have such complex relationships. Members specify their requirements on data exchange using a policy language (CPL). The requirements defined with the use of CPL form the local data exchange policies of members. Local policies are defined separately for data sharing and data acquisition policies. This property allows asymmetric relations on data exchange. For example, a member does not necessarily have to acquire the data that the other members dictate to share. By using these two policies, members specify statements of who to share/acquire and what to share/acquire. The statements are defined using \emph{conditional} and \emph{selection} expressions. Selections allow members to filter data and limit the data to be exchanged, whereas conditional expressions allow members to define logical statements. Another advanced property of CPL is predefined \emph{data-dependent conditionals} for calculating the statistical metrics between member's data. For instance, members can define a conditional to compute the intersection size of data columns without disclosing their data. This allows members to define content-dependent conditional data exchange in their policies. 

Once members have defined their local policies, they negotiate a sharing agreement. The guarantee provided by \Curie is that all data exchanged among members will respect the agreement. The agreement is executed in a multi-party privacy-preserving prediction model enhanced with optional differential privacy guarantees. In this work, we make the following contributions:

\begin{itemize}[noitemsep,topsep=0pt]
\item We introduce \Curie, an approach for secure data exchange among members that have complex relationships. \Curie includes CPL policy language allowing members to define complex specifications of data exchange requirements, negotiate an agreement, and execute agreements in a multi-party predictive model that policies respect the negotiated policy.

\item We validate \Curie through an example of real healthcare application used to prescribe warfarin dosage. A privacy-preserving joint dose model among medical institutions is compiled with the use of various data exchange policies while protecting the privacy of members' healthcare records.

\item We show \Curie incurs low overhead and policies are effective at improving the dose accuracy of medical institutions.
\end{itemize}

\vspace{3pt}

\noindent We begin in the next section by defining the analysis task and outlining the security and attacker models. 

\section{Problem Scope and Attacker Model}
\label{sec:problemScope}
\noindent\textbf{Problem Scope.} We introduce Curie Policy Language (CPL) to express data exchange requirements of distributed members. Unlike the programming languages used for writing secure multiparty computation (MPC)~\cite{henecka2010tasty, rastogi2014wysteria} and the frameworks designed for privacy-preserving machine learning (ML)~\cite{liu2015oblivm, ohrimenko2016oblivious, bogdanov2016rmind, el2013secure, mohassel2017secureml}, CPL is a policy language in a Backus Normal Form (BNF) notation to express the conflicting relationships of members on data sharing. Members can express data exchange requirements using the conditionals, selections, and secure pairwise data-dependent statistics. \Curie then enforces the policy agreements in a shared predictive model through an MPC protocol that ensures members comply with the policies as negotiated. 

We integrate \Curie into 24 medical institutions. Without deployment of \Curie, institutions compute warfarin dosage of a patient using a model computed on their local patient records. \Curie allows institutions to construct various consortia wherein each member defines a data exchange policy for other members via CPL. This enables institutions to acquire the patient records based on regulations as well as the records that they need to improve the accuracy of their dose predictions.  \Curie implements a privacy-preserving dose model through homomorphic encryption (HE) to enforce the policy agreements of the members. We note that a centralized party in HE cannot provide a privacy-preserving model on negotiated data~\cite{van2010impossibility}. However, \Curie implements a novel protocol that allows institutions to perform local computations by aggregating the intermediate results of the dose model. Additionally, \Curie implements an optional differential private (DP) mechanism that allows institutions to perform differentially-private (DP) secure dose model. DP guarantees that no information leaks on the targeted individual (\ie patient) with high confidence from the released dose model.

\vspace{2pt}\noindent\textbf{Threat Model.} We consider a semi-honest adversary model. That is, members in a consortium runs the protocol exactly as specified, yet they try to learn the dataset inputs of the other members as much as possible from their views of the protocol. Additionally, we consider non-adaptive adversary wherein members cannot modify inputs of their dataset once the protocol on shared data is initiated. 

\section{Organizational Data Exchange}
\label{sec:requirements}
\begin{figure}[t!]
\centering
\includegraphics[width=1\columnwidth]{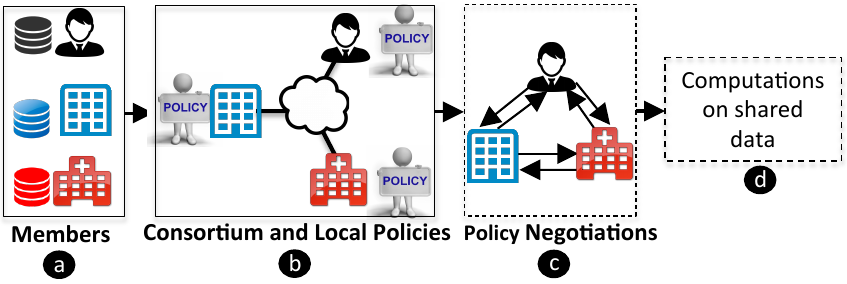}
\caption{\Curie data exchange process in a collaborative learning setting. The dashed boxes show data remains confidential.}
\label{fig:process}
\end{figure}

Depicted in Figure~\ref{fig:process}, \Curie includes two independent parts: policy management and multiparty secure computation.

\vspace{2pt}\noindent\textbf{Policy Management.} We define a \emph{consortium} that is a group made up of two or more members--individuals, companies or governments (\circled{a}). Members of a consortium aim to compute a predictive model $m$ over their confidential data in a secure manner. 
For instance, data may be curated from medical history of patients or financial reports of companies with the objective of building an ML model. Moreover, each member wants to enforce a set of local constraints toward other consortium members to control their requirements on how and with whom they share their confidential data. 
These constraints define a member's interest, trust, regulations and data demands, and also impacts the accuracy of a model $m$. Thus, there is a need for connecting data needs of members to the privacy-preserving models. In \Curie, each member of a consortium defines a \emph{local policy} (\circled{b}). The local policy of a member dictates the requirements of data exchange as follows:
\begin{enumerate}[leftmargin=*]
\item The member wishes to specify with whom to share and acquire data (\emph{partnership requirement}).
\item The member wishes to define what data to share and acquire (\emph{sharing and acquisition requirement}).
\end{enumerate}
In this, the member wishes to refine its sharing and acquisition requirements to express the following:
\begin{enumerate}[leftmargin=*] 
\item The member wishes to dictate a set of conditions to restrict data sharing and select which data to be acquired (\emph{conditional selective share and acquisition}); and 
\item The member wishes to dictate conditionals based on the other member's data (\emph{data-dependent conditionals}).
\end{enumerate}

\begin{figure}[ht!]
\centering
\includegraphics[width=0.6\columnwidth]{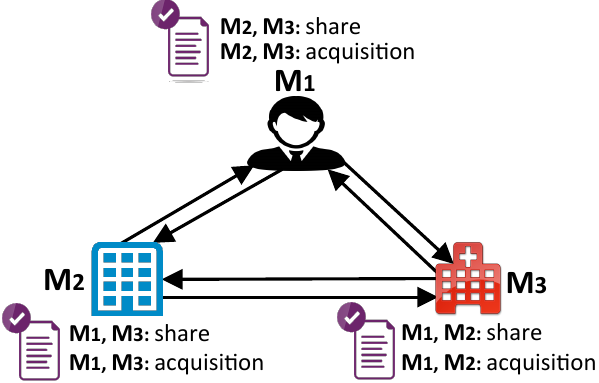}
\caption{An example consortium of three members.}
\label{fig:consortia-members}
\end{figure}

The policy of members need not be-nor are likely to be-symmetric. Local policy is defined with requirements for sharing and acquisition that is tailored to each partner member in the consortium--thus allowing each pairwise sharing to be unique.  Here, the local policies are used to negotiate pairwise sharing within the consortium. To illustrate how members negotiate an agreement, consider the consortium of three members in Figure~\ref{fig:consortia-members}.

Each member initiates pairwise policy negotiations with other members to reconcile contradictions between acquisition and share policies (\circled{c}). A member starts the negotiation by sending a request message including the acquisition policy defined for a member. When a member receives the acquisition policy, it reconciles the received acquisition policy with its share policy specified for that member. Three negotiation outcomes are possible: the acquisition policy is entirely satisfied, partially satisfied with the intersection of acquisition and share policies or is an empty set. A member completes its negotiations after all of its acquisition policies for interested parties are negotiated. 

\vspace{2pt}\noindent\textbf{Computations on Negotiated Data.} Once members negotiate their policies (\circled{d}), \Curie provides a multiparty data exchange device using secure multi-party computation techniques enhanced with (optional) differential privacy guarantees. This device ensures data and individual privacy. The guarantee provided by \Curie is that all computations among members will respect their policies. 

To ensure data privacy, \Curie includes cryptographic primitives such as Homomorphic Encryption (HE) and garbled circuits from the secure multi-party computation literature that allows members to perform computations on negotiated data with no disclosed data from any single member. At the end of the secure computation, all of the parties obtain a final predictive model based on their policy negotiations. To ensure the privacy of the individuals in the dataset, which the final model is computed on, \Curie integrates Differential Privacy (DP). DP protects against an attacker who tries to extract a particular individual's data in the dataset from the final computed model at the end of the secure computation protocol.

\section{Curie Policy Description Language}
\label{sec:policy-lan}
We now illustrate the format and semantics of the \Curie Policy Language (CPL). A BNF description of CPL is presented in Appendix~\ref{sec:CPL}. Turning to the example consortium in Figure~\ref{fig:consortia-members} established with three members, each member defines its requirements for other members on a dataset having the columns of age, race, genotype, and weight (see Table~\ref{table:example-data-exchange}). The criteria defined by members are used throughout to construct their local policies.

\vspace{2pt}\noindent\textbf{Share and Acquisition Clauses.} \Curie policies are collections of clauses. The collection of clauses for partners defines the local policy of a member. The clauses allow each member to dictate a member specific policy for each other member. Clauses have the following structure:

\vspace{-2pt}
\begin{center}
$\langle \textmd{clause tag} \rangle : \langle \textmd{members}\rangle : \langle \textmd{conditionals}\rangle :: \langle \textmd{selections}\rangle$;
\end{center}
\vspace{-2pt}

\noindent Clause tags are reference names for policy entries. \textit{Share} and \textit{acquire} are two reserved tags. Those clauses are comprised of three parts. The first part, \textit{members}, defines a list of members with whom to share and acquire. This can be a single member or a comma-separated list of members. An empty member entry matches all members. The second part, \textit{conditionals}, is a list of conditions controlling when this clause will be executed. A condition is a Boolean function which expresses whether the share or acquire is allowed or not. For instance, a member may define a condition where the data size is greater than a specific value. Only if all conditions listed in conditionals are true, then this clause is executed. Last part, \textit{selections}, states what to share or acquire. It can be a list of filters on a member's data. For instance, a member may define a filter on a column of a dataset to limit acquisition to a subset of the dataset. More complex selections can be assigned using member defined sub-clauses. A sub-clause has the following structure:
\vspace{-2pt}
\begin{center}
$\langle \textmd{tag} \rangle : \langle \textmd{conditionals}\rangle :: \langle \textmd{selections}\rangle ; $
\end{center}
\vspace{-2pt}
\noindent where \textit{tag} is the name of sub-clause; conditionals is, as explained above, a list of conditions stating whether this clause will be executed; selections is a list of filters or a reference to a new sub-clause. Complex data selection can be addressed with nested sub-clauses.  

CPL allows members to define multiple clauses. For instance, a member may share a distinct subset of data for different conditions. CPL evaluates multiple clauses in a top-down order. When conditionals of a clause evaluate to false, it moves to the next clause until a clause is matched or it reaches end of the policy file. 
\begin{table}[t!]
\centering
\setlength{\tabcolsep}{3pt} 
\renewcommand{\arraystretch}{0.8}
\resizebox{\columnwidth}{!}{%
{\small{
\begin{tabular}{|p{0.8cm} p{8cm}|}
\hline
\multicolumn{2}{|c|}{\textbf{Consortia member: \textrm{M$_1$}}} \\
\textrm{M}$_2$\textendash & desires to acquire complete data of users who are older than 25 \\
\textrm{M}$_2$\textendash & shares its complete data \\
\textrm{M}$_3$\textendash & desires to acquire Asian users such that the Jaccard similarity of its age column and \textrm{M}$_3$'s age column is greater than 0.3 \\
\textrm{M}$_3$\textendash & shares its complete data \\\hline
\multicolumn{2}{|c|}{\textbf{Consortia member: \textrm{M$_2$}}} \\  
\textrm{M}$_1$\textendash & desires to acquire complete data \\
\textrm{M}$_1$\textendash & limits its share to EU and NATO citizen users if \textrm{M}$_1$ is both NATO and EU member and located in North America. Otherwise, it shares only White users \\
\textrm{M}$_3$\textendash & desires to acquire complete data if \textrm{M}$_3$ is a NATO member \\
\textrm{M}$_3$\textendash & shares its complete data \\\hline
\multicolumn{2}{|c|}{\textbf{Consortia member: \textrm{M$_3$}}} \\
\textrm{M}$_1$\textendash & desires to acquire complete data of users having genotype `A/A' \\
\textrm{M}$_1$\textendash & share complete data if intersection size of its and \textrm{M}$_1$'s genotype column is less than 10. Otherwise, it shares data of users that weigh more than 100 pounds \\
\textrm{M}$_2$\textendash & desires to acquire complete data \\
\textrm{M}$_2$\textendash & shares complete data if \textrm{M}$_2$ is EU member and its data size is greater than 1K \\ \hline
\end{tabular}
}}}
\caption{An example of member's data exchange requirements.}
\label{table:example-data-exchange}
\end{table}

\vspace{2pt}\noindent\textbf{Conditionals and Selections.} We present the use of conditionals and selections through policies with examples. Their format and semantics are detailed. Consider an example of two members, M$_1$ and M$_2$, within a consortium. They define their local policies as:
\vspace{-2pt}
\textsf{
{\small{
\begin{align*}
  @\textsf{M}_1 ~~&\textsf{acquire :  M$_2$ : ~ :: s$_1$ ;}  \\[-1ex]
  & \textsf{share :  M$_2$  : ~ :: ~ ;} \\[-1ex]
  @\textsf{M}_2 ~~ & \textsf{acquire :  M$_1$ : ~ :: ~ ; } \\[-1ex]
  &\textsf{share :  M$_1$  :  c$_1$, c$_2$ :: fine-select ;}  \\[-1ex]
  &\textsf{fine-select    :  c$_3$ ::  s$_2$ ;}  \\[-1ex]
  &\textsf{fine-select    : ~  ::  s$_3$ }; 
\end{align*}
}}}
\vspace{-2pt}
\noindent where \textsf{c$_1$}, \textsf{c$_2$} and \textsf{c$_3$} are conditionals,  \textsf{s$_1$}, \textsf{s$_2$} and \textsf{s$_3$} are selections and \textsf{fine-select} is a tag defined by \textsf{M}$_2$. 

The acquire clause of \textsf{M$_1$} states that data is requested from \textsf{M$_2$} after it applies \textsf{s$_1$} selection (\eg \textsf{age $>25$}) to its data. In contrast, its share clause allows complete share of its data if \textsf{M$_2$} requests. On the other hand, the acquisition clause of \textsf{M$_1$} dictates requesting complete data from \textsf{M$_2$}. However, \textsf{M$_2$} allows data sharing if the acquisition clause issued by \textsf{M$_1$} holds \textsf{c$_1$} $\land$ \textsf{c$_2$} conditions (\eg is both NATO and EU member). Then, \textsf{M$_2$} delegates selection to member-defined \textsf{fine-select} sub-clauses. \textsf{fine-select} states that if the request satisfies the \textsf{c$_3$} condition (located in North America) then the request is met with the data that is selected by the \textsf{s$_2$} selection (\eg limits share of its data to NATO and EU member country citizens). Otherwise, it shares data that is specified by selection \textsf{s$_3$} (White users).

CPL supports selections through filters. A filter contains zero or more operations over data inputs describing the share and acquisition criteria to be enforced. Operations are defined as keywords or symbols such as $<$, $>$, $=$, $in$, $like$, and so on. Selections and filters are defined in CPL as follows:
\vspace{-2pt}
\setlength{\grammarparsep}{2pt plus 1pt minus 1pt} %
\setlength{\grammarindent}{8em} 
\renewcommand{\syntleft}{$\langle$\normalfont \mdseries} 
\textsf{
\begin{grammar}
<selections> ::= <filters> |  <tag> \\
<filters> ::= <filter> [`,' <filters>] \\
<filter> ::= <var> <operation> <value> | `'
\end{grammar}}
\vspace{-2pt}
\noindent Selections are executed when conditionals evaluated to be true. Conditionals can be consortium and dataset-specific. For instance, a member may require other members to be in a particular country or to be in an alliance such as NATO and to have their dataset size greater than a particular value. Such conditionals do not require any data exchange between members to be evaluated. However, members may want to incorporate a relation between their data and other member's data into their policies as detailed next. 

\vspace{2pt}\noindent\textbf{Data-dependent Conditionals.} A member's decision on whether to share or to acquire data can depend on other member's data. Simply put, one example of a data-dependent conditional among two members could be whether the intersection size of the two sets (\eg a specific column of a dataset) is not too high. Considering such knowledge, a member can make a conditional decision about share or acquisition of that data. For instance, consider a list of private IP addresses used for blacklisting the domains. If a member knows that the intersection size is close to zero, then the member may dictate an acquire clause to request complete features from that member based on IP addresses ~\cite{freudiger2015controlled}. 

CPL defines an \textsf{evaluate} keyword for data-dependent conditionals through functions on data. Data-dependent conditionals take the following form:
\vspace{-2pt}
\setlength{\grammarparsep}{2pt plus 1pt minus 1pt} %
\setlength{\grammarindent}{8em} 
\renewcommand{\syntleft}{$\langle$\normalfont \mdseries} 
\textsf{
\begin{grammar}
<conditionals> ::= <var>`='<value> [`,' <conditionals>]
\alt `evaluate' `(' <data_ref> `,' <alg_arg> `,'
<thshold_arg> `)'  [`,' <conditionals>] | `'
\end{grammar}
}
\vspace{-2pt}
A member that uses the data-dependent conditionals defines a reference data (\textsf{data_ref}) required for a such computation, an algorithm (\textsf{alg_arg}) and a threshold (\textsf{thshold_arg}) that is compared with the output of the computation. CPL includes four algorithms for data-dependent conditionals (see Table~\ref{table:policy-functions}). To be brief, intersection size measures the size of the overlap between two sets; Jaccard index is a statistic measure of similarity between sets; Pearson correlation is a statistical measure of how much two sets are linearly dependent; and 
Cosine similarity is a measure of similarity between two vectors. Each algorithm is based on a different assumption about the underlying reference data. However, central to all of them is to privately (without leaking any sensitive data) measure a relation between two members' data to offer an effective data exchange. We note that these algorithms are found to be effective in capturing input relations in datasets~\cite{freudiger2015controlled, garrido2016shall}.

\begin{table}[t!]
\centering
\renewcommand{\arraystretch}{1.2}
\setlength{\tabcolsep}{0.8pt}
\resizebox{\columnwidth}{!}{%
{\small{
\begin{tabular}{|l|c|c|c|}
\hline
\textbf{Pairwise alg.} & \textbf{Output} & \textbf{Private protocol}    & \textbf{Proof} \\ \hline\hline
\textrm{Intersection size}  & $|\mathcal{D}_i \cap \mathcal{D}_j|$ &  Intersection cardinality & \cite{de2012fast} \\\hline
\textrm{Jaccard index}  & $ (|\mathcal{D}_i \cap D_j|)/ (|\mathcal{D}_i \cup \mathcal{D}_j|)$ & Jaccard similarity  &   \cite{blundo2013espresso}      \\\hline
\textrm{Pearson correlation}    & $(COV(\mathcal{D}_i, \mathcal{D}_j))/(\sigma_{\mathcal{D}_i} \sigma_{\mathcal{D}_j})$ & Garbled circuits & \cite{huang2011faster}     \\\hline
\textrm{Cosine similarity}   & $(\mathcal{D}_i \mathcal{D}_j)/(\|{\mathcal{D}_i}\| \|{\mathcal{D}_j}\|)$& Garbled circuits & \cite{huang2011faster}\\ \hline
\end{tabular}}}
}
\caption{CPL data-dependent conditional algorithms. Two members of a consortium use the conditionals to compute the pairwise statistics. The members then use the output of the algorithm to determine whether to acquire or share data from another party. ($\mathcal{D}_i$ and $\mathcal{D}_j$ are the inputs of a dataset, and $\sigma$ is std. deviation).}
\label{table:policy-functions}
\end{table}

Data-dependent conditionals are implemented through private protocols (as defined in Table~\ref{table:policy-functions}). These protocols are implemented with the cryptographic tools of garbled circuits and private functions. Protocols preserve the confidentiality of data. That is, each member gets the output indicated in Table~\ref{table:policy-functions} without revealing their sensitive data in plain text. After the private protocol terminates, the output of the algorithm is compared with a threshold value set by the requester. If the output is below the threshold value, the conditional is evaluated to true. Turning to above example \textsf{M$_3$} joins the consortium. \textsf{M$_1$} and  \textsf{M$_2$} extend their local policies for \textsf{M$_3$}:

\vspace{-2pt}
{\small{
\textsf{
\begin{align*}
  @\textsf{M}_1~~&\textsf{acquire :  M$_3$ : } \textsf{evaluate(local data,} \textsf{ 'Jaccard', 0.3)  :: race=Asian;} \nonumber \\[-1ex]
   & \textsf{share :  M$_3$ : ~ :: ~ ; }\\[-1ex]
   @\textsf{M}_2~~& \textsf{acquire :  M$_3$ : M$_3$ in \$NATO :: ~ ; } \\[-1ex]
   & \textsf{share :  M$_3$ : ~ :: ~ ; }\\[-1ex]
  @\textsf{M}_3~~& \textsf{acquire :  M$_1$ : ~ :: Genotype = 'A/A' ; } \\[-1ex]
   &\textsf{share :  M$_1$  : }  \textsf{evaluate(local data,} \textsf{'intersection size', 10) :: ~ ;}\\[-1ex]
   &\textsf{share :  M$_1$  : ~ :: weight$>$150 ;}   \\[-1ex]
   & \textsf{acquire :  M$_2$ : ~ :: ~ ; } \\[-1ex]
   & \textsf{share :  M$_2$ : M$_2$ in \$EU, size(data)$>$ 1K :: ~ ; }
\end{align*}
}}}
\vspace{-2pt}

\noindent The acquire clause of \textsf{M$_1$} defines a data-dependent conditional for \textsf{M$_3$}. It defines a Jaccard measure on its local data through \textsf{evaluate} keyword and sets its threshold value equal to 0.3. \textsf{M$_3$} agrees to share its local data with \textsf{M$_1$} if \textsf{intersection size} of its local data is less then 10. Otherwise, it consults the next share clause defined for {M$_1$} which states that an individual's weight greater than 150 pounds will be shared. All other share and acquire clauses are trivial. Members agree to share and acquire complete data based on data size (data size $>$ 1K), alliance membership (\eg NATO or EU member) and inputs (\eg genotype).

Putting pieces together, CPL allows members independently define a data exchange policy with share and acquire clauses. The policies are dictated through conditionals and selections. This allows members to dictate policies in complex and asymmetric relationships. Defined in Section~\ref{sec:requirements}, CPL provides members to dictate partnership, share, acquisition, and data-dependent conditionals.

\begin{table*}[h!]
\centering
\def\arraystretch{1}
\setlength{\tabcolsep}{1pt} 
\resizebox{\textwidth}{!}{%
{\small{
\begin{tabular}{@{}|cclcc|}
\hline
\textbf{\begin{tabular}[c]{@{}c@{}}Policy ID\end{tabular}} & \textbf{\begin{tabular}[c]{@{}c@{}}Consortium Name\end{tabular}} & \multicolumn{1}{c}{\textbf{Policy Definition}}& \textbf{\begin{tabular}[c]{@{}c@{}}Acquisition Policy\end{tabular}} & \textbf{\begin{tabular}[c]{@{}c@{}}Share Policy\end{tabular}} \\ \hline \hline

\rowcolor[HTML]{DCDCDC} 
P.1                                                          & Single Source                                                  & \begin{tabular}[c]{@{}l@{}}Each member uses its local patient dataset to learn warfarin dose model. \end{tabular}                                                               & \xmark                                                                  & \xmark                                                            \\ \hline
P.2                                                        & Nation-wide                                                    & \begin{tabular}[c]{@{}l@{}}Members in the same country establish a consortium based on state and country laws. \end{tabular}                  & \cmark                                                                  & \cmark                                                            \\ \hline

\rowcolor[HTML]{DCDCDC} 
P.3                                                          & Regional                                                       & \begin{tabular}[c]{@{}l@{}}Members in the same continent establish a consortium. \end{tabular} & \cmark                                                                  & \cmark                                                            \\ \hline

P.4                                                         & NATO-EU                                                        & \begin{tabular}[c]{@{}l@{}}NATO and EU members establish a consortium independently based on their mutual agreements.\end{tabular}                                  & \cmark                                                                  & \cmark                                                            \\ \hline

\rowcolor[HTML]{DCDCDC} 
P.5                                                  & Global                                              & \begin{tabular}[c]{@{}l@{}}Members exchange their complete data to build the warfarin dose model. \end{tabular}                                                                     & \cmark                                                                  & \cmark     \\ \hline

\end{tabular}}}
}
\caption{Consortia constructed among members. Acquisition and share policies of members for each consortium are studied in Section~\ref{sec:evaluation}.}
\label{table:policies}
\end{table*}

\vspace{2pt}\noindent\textbf{Policy Negotiation and Conflicts.} Data exchange between members is governed by matching share and acquire clauses in each member's respective policies. Both share and acquire clauses state conditions and selections on the data exchanged. Consider two example local policies with a share clause $@m_2$ ($share : m_1:c_1::s_1$) and matching acquire clause $@m_1$ ($acquire: m_2:c_2:s_2$). \Curie's negotiation algorithm respects both autonomy of the data owner and the needs of the requester. It conservatively negotiates share and acquire clauses such that it will return the \emph{intersection} of respective data sets in resulting policy assignment. The resolved policy in this example is $share:m_1:c_1 \land c_2 :: s_1 \land s_2$ which states that the data exchange from $m_2$ to $m_1$ is subject to both $c_1$ and $c_2$ conditionals and resulting sharing has $s_1$ and $s_2$ selections on $m_2$'s data. This authoritative negotiation makes sure no member's data is shared beyond its explicit intent, regardless how the other members' policies are defined. This is because negotiation fulfilling the criteria for each clause is based on the union of logical expressions defined in two policies. Each member runs the negotiation algorithm for members found in their member list. After all members terminate their negotiations, the negotiated policy is enforced in computations.

\section{Deployment of Curie} 
To validate \Curie in a real application, we integrated \Curie into 24 medical institutions. Each institution wants to compute a warfarin dose model on the distributed dataset without disclosing the patient health-care records. Without deployment of \Curie, institutions compute warfarin dosage of a patient using a model computed on their local patient data. \Curie first enables institutions to negotiate their data exchange requirements through CPL. In this, \Curie allows members to construct various consortia wherein each member defines a data exchange policy for other members. The next step is to compute a privacy-preserving dose model such that each party does not learn any information about the patient's records of other medical institutions and respects the policy negotiated. \Curie implements a secure dose protocol through homomorphic encryption (HE) to enforce the policy agreements of the members. We next present the deployment of \Curie to institutions (Section~\ref{sec:deploymentSetup}) and integration of policy agreements in warfarin dose model (Section~\ref{sec:secure-dose-algorithm}).

\subsection{Deployment Setup}
\label{sec:deploymentSetup}
\noindent\textbf{Warfarin-} known as the brand name Coumadin is a widely prescribed  (over 20 million times each year in the United States) anticoagulant medication. It is mainly used to treat (or prevent) blood clots (thrombosis) in veins or arteries. Taking high-dose warfarin causes thin blood which may result in intracranial and extracranial bleeding. Taking low doses causes thick blood which may result in embolism and stroke. Current clinical practices suggest a \emph{fixed} initial dose of 5 or 10 mg/day. Patients regularly have a blood test to check how long it takes for blood to clot (international normalized ratio (INR)). Based on the INR, subsequent doses are adjusted to maintain the patient's INR at the desired level. Therefore, it is important to predict the proper warfarin dose for the patients.

\vspace{2pt}\noindent\textbf{Consortium Members.} 24 medical institutions from nine countries and four continents individually collected the largest patient data for predicting \emph{personalized} warfarin dose (see Appendix~\ref{sec:appendix-members} for details of members involved in the study). Members collect 68 inputs from patients' genotypic, demographic, background information, yet a long study concluded that eight inputs are sufficient for proper prescriptions~\cite{international2009estimation}. 

\vspace{2pt}\noindent\textbf{Warfarin Dose Prediction Model.} To determine the proper personalized warfarin dosage, a long line of work concluded with an algorithm of an ordinary linear regression model~\cite{international2009estimation}. The model is a function $f : \mathcal{X} \to \mathcal{Y}$ that aim at predicting targets of warfarin dose $y \in \mathcal{Y}$ given a set of patient inputs $x \in \mathcal{X}$. We represent the patient dataset of each member  $\mathcal{D}_i = \{(x_i, y_i)\}_{i=1}^n$, and a loss function $\ell : \mathcal{Y} \times \mathcal{Y} \to [0, \infty)$. The loss function penalizes deviations between true dose and predictions. Learning is then searching for a dose model $f$ minimizing the average loss:

{\small{
\begin{equation}
  \mathcal{L}(\mathcal{D}, f) =  \frac{1}{n} \sum_{i=1}^n \ell(f(x_i), y_i).
  \label{eq:dose-algorithm}
\end{equation}}}

The dose model reduces to minimizing the average loss $\mathcal{L}(\mathcal{D}, f)$ with respect to the parameters of the model $f$. The model is linear, \ie $f(x) = \alpha^\top x + \beta$, and the loss function is the squared loss $\ell(f(x),y) = (f(x) - y )^2$. The dose model gives as well or better results than other more complex numerical methods and outperforms fixed-dose approach\footnote{The model has been released online \url{http://www.warfarindosing.org} to help doctors and other clinicians for predicting ideal dose of warfarin.}~\cite{international2009estimation}. We re-implemented the algorithm in Python by direct translation from the authors' implementation and found that the accuracy of our implementation has no statistically significant difference.

\begin{figure}[t!]
\centering
\includegraphics[width=1\columnwidth]{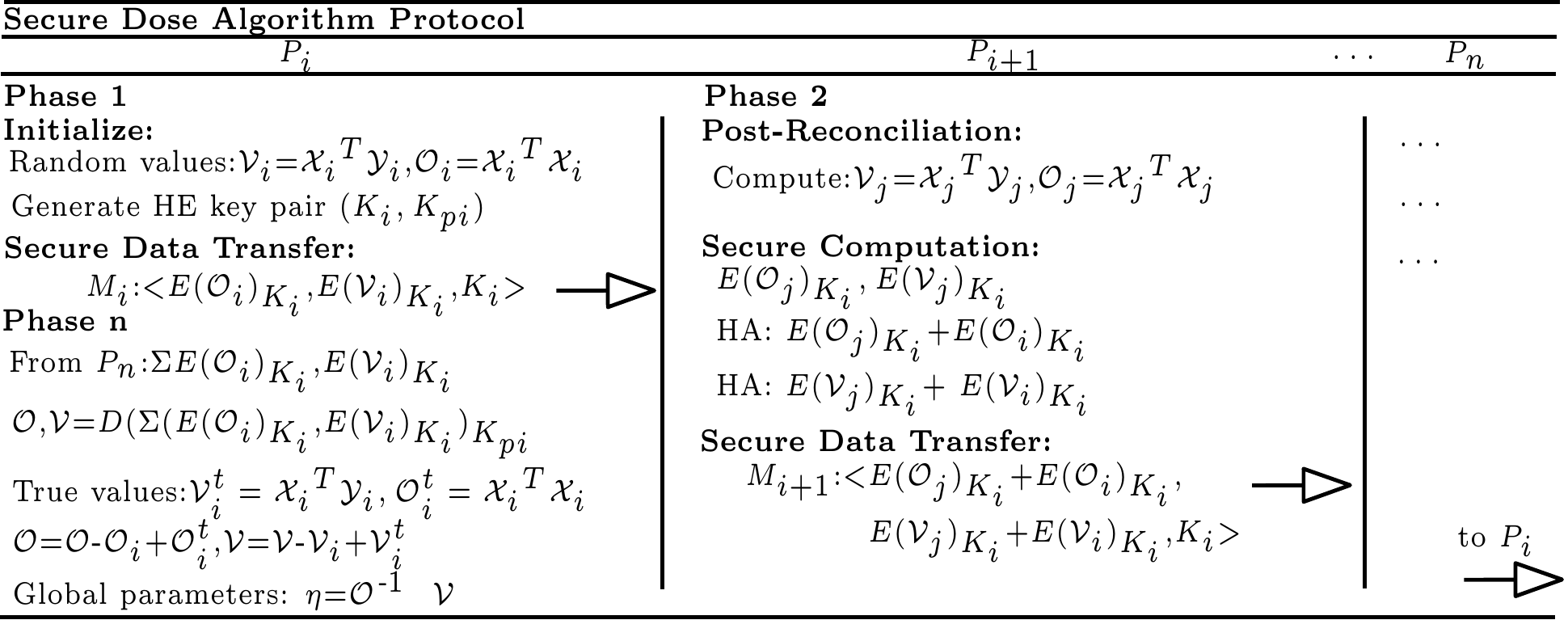}
\caption{Secure dose algorithm protocol: Member ($P_i$) starts the protocol, the procedures and message flow among members are highlighted in boldface. At the final phase, $P_i$ is able to compute the dose model coefficients from the negotiated data.}
\label{fig:secure-dose-algorithm}
\end{figure}

\vspace{2pt}\noindent\textbf{Consortia and Member Policies.} We define consortia among medical institutions that they state partnerships for data exchange. Table~\ref{table:policies} summarizes the consortia. The consortia are defined based on statute and regulations between members, as well as regional, and national partnerships are studied based on their countries~\cite{nato-share,arra, hitech, eu-data-share}. For example, NATO allied medical support doctrine allows strategic relationships that are otherwise not obtainable by non-NATO members. Each member in a consortium exchanges data with other members based on its CPL policy. Various acquisition and share policies of CPL are studied via conditionals and selections in Section~\ref{sec:evaluation}. We note that policy construction is a subjective enterprise. Depending on the nature and constraints of a given environment, any number of policies are appropriate. Such is the promise of policy defined behavior; alternate interpretations leading to other application requirements can be addressed through CPL.

\subsection{Privacy-preserving Dose Prediction Model} 
\label{sec:secure-dose-algorithm}
The computation of \emph{local dose} model of a medical institution is straightforward: a member calculates the dose model through Equation~\ref{dose-algorithm-matrix} with the use of patient data collected locally. To implement a privacy-preserving dose model among consortia members of medical institutions, we define the dose prediction formula stated in Equation~\ref{eq:dose-algorithm} in a matrix form by minimizing with maximum likelihood estimation:

{\small{
\begin{equation}
\vspace{-3pt}
\beta = \inv{(\mathcal{X}^\intercal \mathcal{X})} {\mathcal{X}}^\intercal \mathcal{Y}, 
\label{dose-algorithm-matrix}
\vspace{-3pt}
\end{equation}}}

\noindent where $\mathcal{X}$ is the input matrix, $\mathcal{Y}$ is the dose matrix, and $\beta$ is the coefficients of the dose model. 

\Curie allows members to collaboratively learn a dose model without disclosing their patient records and guarantees data sharing complies with the policy as negotiated. As
shown in Equation~\ref{eq:pooled-dose}, each member translates its negotiated data into neutral input matrices~\cite{wu2011wireless}. 
Particularly, patient samples to be exchanged by each member are computed as an input matrix $\mathcal{X}_{0},\ldots, \mathcal{X}_{n}$ and dose matrix $\mathcal{Y}_{0}, \ldots, \mathcal{Y}_{n}$. The transformation defines each member's \emph{local statistics} $\mathcal{O}_{i}= \mathcal{X}^\intercal \mathcal{X}$ and $\mathcal{V}_{i}=\mathcal{X}^\intercal \mathcal{Y}$. Local statistics is the output of the negotiation of each member in a consortium. The aggregation of the local statistics corresponds to a \emph{negotiated dataset} which is the exact amount that a member negotiates to obtain from other members in a consortium. \Curie constructs the dose algorithm of the negotiated dataset as a concatenation of members' local statistics as follows:

{\small{
\vspace{-3pt}
\begin{align*}
\mathcal{X}^\intercal \mathcal{X} =& \Big[\mathcal{X}_1^\intercal | \ldots | \mathcal{X}_n^\intercal \Big]  {\Big[\mathcal{X}_1 | \ldots | \mathcal{X}_n\Big]}^\intercal = \sum_{i=1}^n \mathcal{X}_i^\intercal \mathcal{X}_i = \sum_{i=1}^n\mathcal{V}_i = \mathcal{V}\\[-0.3em]
\mathcal{X}^\intercal \mathcal{Y} =& \Big[\mathcal{X}_1^\intercal | \ldots | \mathcal{X}_n^\intercal\Big]  {\Big[\mathcal{Y}_1 | \ldots | \mathcal{X}_n\Big]}^\intercal = \sum_{i=1}^n \mathcal{X}_i^\intercal \mathcal{Y}_i = \sum_{i=1}^n \mathcal{O}_i = \mathcal{O} \numberthis 
\label{eq:pooled-dose}
\vspace{-3pt}
\end{align*}}}

In Equation~\ref{eq:pooled-dose}, a member computes model coefficients using the sum of other members local statistics. The local statistics includes $m \times m$ constant matrices where $m$ is the number inputs (independent of number of dataset size). Using this observation, a party computes the coefficients of the negotiated dataset:

{\small{
\vspace{-3pt}
\begin{equation}
{\eta}^{(negotiated)}= \inv{(\mathcal{X}^\intercal \mathcal{X})} \mathcal{X}^\intercal \mathcal{Y} = \inv{\mathcal{O}}\mathcal{V}
\vspace{-3pt}
\label{eq:dose-algorithm-matrix}
\end{equation}}}

\noindent In Equation~\ref{eq:dose-algorithm-matrix}, while the accuracy objective of the dose model is guaranteed using the coefficients obtained from the sum of local statistics, the exchange of clear statistics among parties may leak information about members' data. A member can infer knowledge about the distribution of each input of other members from matrices of $\mathcal{O}_{i}$ and $\mathcal{V}_{i}$~\cite{el2013secure}. Furthermore, an adversary may sniff data traffic to control and modify exchanged messages. To solve these problems, we use homomorphic encryption (HE) that allows computation on ciphertexts~\cite{DBLP:journals/corr/AcarAUC17}. HE allows members to perform the computation of joint of function without requiring additional communication complexity other than the data exchange. We note that HE itself cannot preserve the confidentiality of data from multiple parties in centralized settings~\cite{VanDijk}. However, \Curie implements a distributed privacy-preserving multi-party dose model, as shown in Figure~\ref{fig:secure-dose-algorithm}.

\begin{figure*}[t!]
    \centering
    \begin{minipage}{0.47\textwidth}
        \centering
        \includegraphics[width=0.72\textwidth]{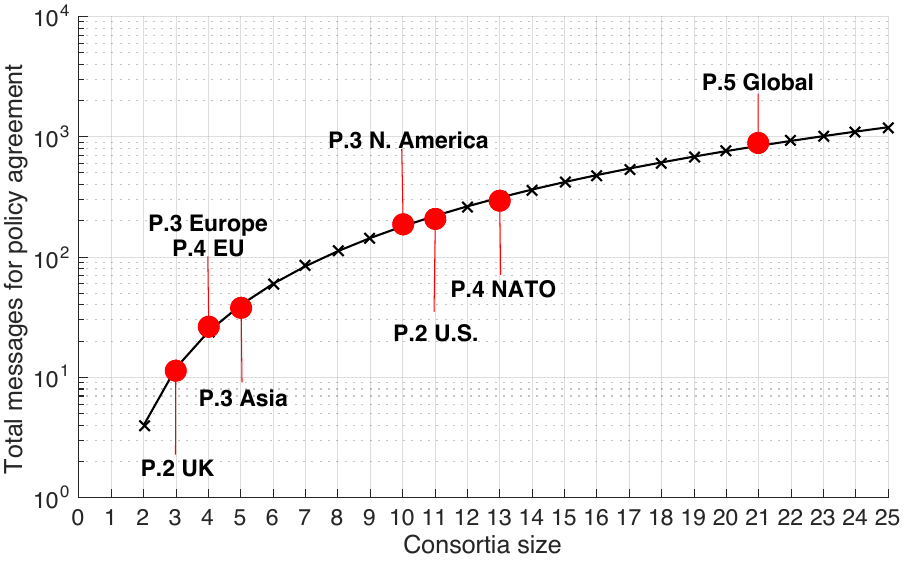} 
        \caption{CPL negotiation cost - Costs associated with a number of varying members in a consortium. Each member defines asymmetric share and acquisition policy for other members. The number of members in warfarin consortia is marked with red circles.}
        \label{fig:members-vs-messages}
    \end{minipage}\hfill
    \begin{minipage}{0.47\textwidth}
        \centering
        \includegraphics[width=0.72\textwidth]{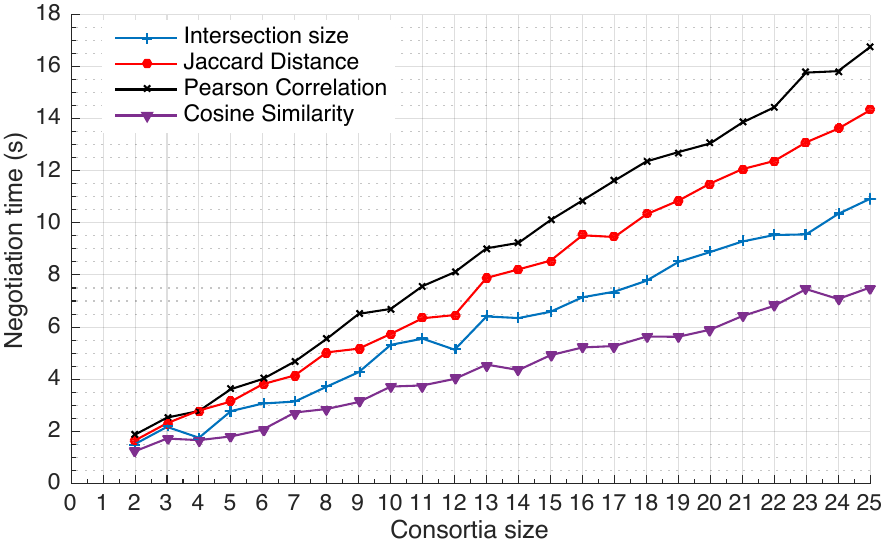} 
        \caption{CPL selections and data-dependent conditional costs - Costs associated with varying members and algorithms. All consortia members agree on policy including a different data-dependent conditional and selections over one input of having 200 samples.}
        \label{fig:members-vs-time}
    \end{minipage}
    \vspace{-10pt}
\end{figure*}

To illustrate, we consider an example session of $n$ members authorized for data exchange in a consortium. In this example, a ring topology is used for secure group communication (\ie $P_i$ talks to $P_{i+1}$, and similarly $P_n$ talks to $P_i$). $P_1$ initially generates a pair of encryption keys using the homomorphic cryptosystem and broadcasts the public key to the members in its member list. $P_1$ then generates random $\mathcal{V}_i$, $\mathcal{O}_i$ and encrypts them $E(\mathcal{O}_i)_{K_{i}}$ and $E(\mathcal{V}_i)_{K_{i}}$ using its public key $K_i$. It starts the session by sending them to the next member in the ring. When next member receives the encrypted message, it adds its local $\mathcal{V}_i$ and $\mathcal{O}_i$ matrices through homomorphic addition to the output of its policy reconciliation for $P_1$ and passes to the next member. Remaining members take the similar steps. Secure computation executes one round per member in which the computation for the particular member visits other members. This allows \Curie to enforce  HE on shared data of a particular member in each round uses and does not suffer insecurities associated with centralized HE constructions~\cite{VanDijk}. 

At the final stage of the protocol, $P_1$ receives the sum statistics of $\mathcal{O}_i$ and $\mathcal{V}_i$ from $P_n$. $P_1$ decrypts the sum of the statistics using its private key and then subtracts the initial random values of $\mathcal{V}_i$, $\mathcal{O}_i$ and adds its true values used for computation of the local dose model coefficients. The final result $\mathcal{O}$ and $\mathcal{V}$ is the coefficients of the dose model that respects $P_1$'s policy negotiations. Other consortium members similarly start the protocol and compute the coefficients. We present the security analysis of the dose protocol in Appendix~\ref{sec:securityAnalysis}, and show its differentially-private extension in Appendix~\ref{sec:dif-privacy}.

\begin{figure*}[th!]
\vspace{6pt}
\centering
\includegraphics[width=1\textwidth]{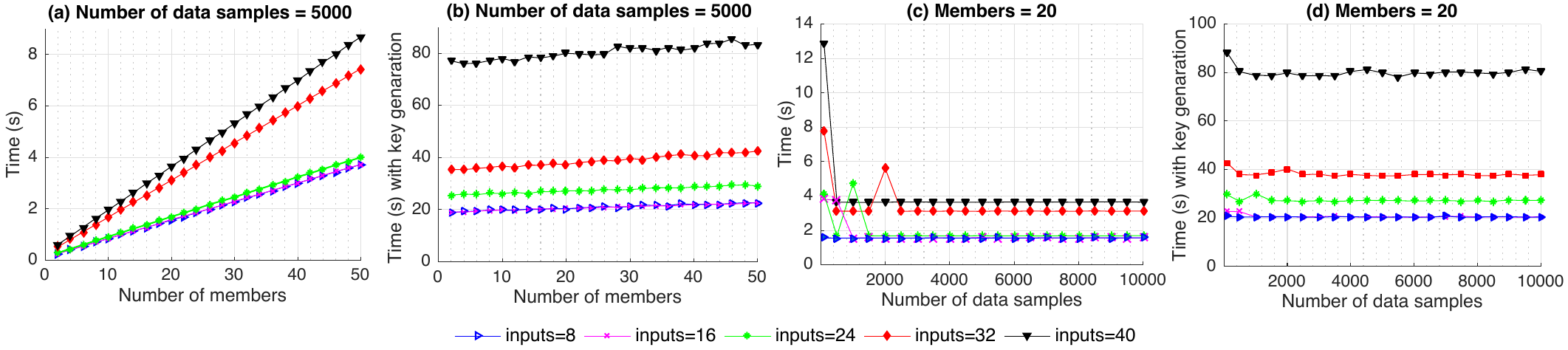}
\caption{CPL performance on privacy-preserving and differential private protocol - All members define an asymmetric share and acquisition policy through selections and conditionals. The agreements of CPL policies between consortia members are studied with the different number of consortia members, data samples, and input size. (Std. dev. of ten runs is $\pm$ 3.6 and $\pm$ 0.3 sec. with and without homomorphic key generation.)}
\label{fig:sec-comp-layer}
\end{figure*}

\section{Evaluation}
\label{sec:evaluation}
This section details the operation of the \Curie through policies. We show how flexible data exchange policies are implemented and operated. We focus on the following questions:
\begin{enumerate}[noitemsep,topsep=3pt]
\item What are the performance trade-offs in configuring CPL?
\item Can members reliably use \Curie to integrate various policies?
\item Do members improve the accuracy of dose predictions with the use of CPL?
\end{enumerate}

The answers to the first two questions are addressed in Section~\ref{sec:comp-results}, and the last question is answered in Section~\ref{sec:policy-results}. As detailed throughout, \Curie allows 50 members to compute the privacy-preserving model using 5K data samples with 40 inputs in less than a minute. We also show how an algorithm with flexible data exchange policies can improve--often substantially--the accuracy of the warfarin dose model accuracy.

\vspace{2pt}\noindent\textbf{Experimental Setup.} The experiments were performed on a cluster of machines with 32 GB of maximum memory and 16-core Intel Xeon CPU at 1.90 GHz, where we use one core to get a lower bound estimate. Each member is simulated in a server that stores its data. Secure computation protocols of \Curie are implemented using the open-source HElib library~\cite{helib}. We set the security parameter of HElib as 128 bits. Multiplication level is optimized per member to increase the number of allowed homomorphic operations without decryption failure and to reduce the computation time. 

We validate the accuracy of dose model in various consortia defined in Table~\ref{table:policies} with members defining different data exchange policies. The dataset used in our experiments contains 5700 patient records from 21 members. 
Dose model accuracy of each member is validated with Mean Absolute Percentage Error (MAPE). MAPE measures the percentage of how far predicted dosages are away from true dosage. Lower values indicate better quality of treatment.

\subsection{Performance Evaluation}
\label{sec:comp-results}
We present the costs associated with various \Curie mechanisms. We illustrate the cost of the CPL in policy negotiations, in the use of data-dependent conditionals, and in the dose algorithm. 

\vspace{-5pt}\subsubsection{CPL Benchmarks} Our first set of experiments characterize the policy construction and negotiation costs. Various consortia and policies are instrumented to analyze the overhead of the number of messages and time required to compute the CPL selections and data-dependent conditionals. All the costs not specific to the policies are excluded in measurements (\eg network latency). The benchmark results are summarized in Figure~\ref{fig:members-vs-messages} and \ref{fig:members-vs-time} and discussed below.

\begin{figure*}[t!]
\centering
\includegraphics[width=1\textwidth]{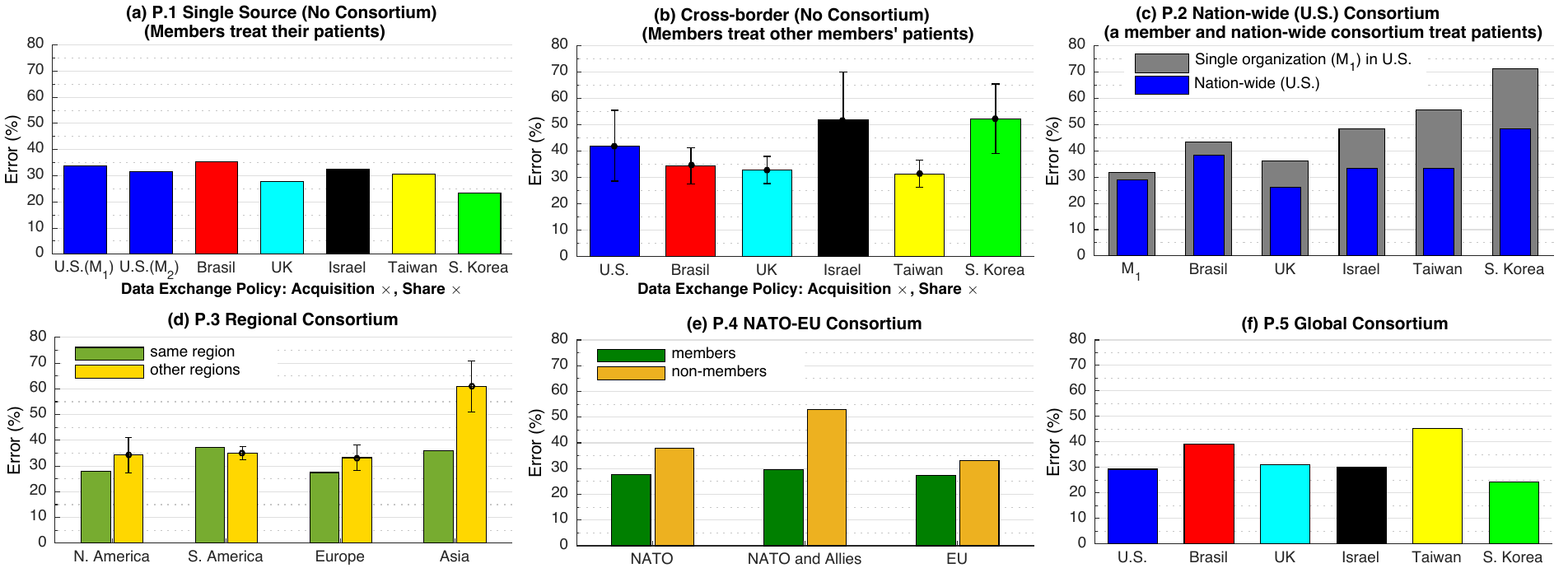}
\caption{The implication of policies on model accuracy - errors are validated in various consortia through data exchange policies. Figure 6(c-f): The local acquisition policies of members comply with the sharing policy within a consortium (\ie members acquire complete data of the consortia members. Std. devs. of errors are within \%5, if not illustrated).}
\label{fig:policy-results}
\end{figure*}

Figure~\ref{fig:members-vs-messages} shows the number of messages for policy construction required for different consortia size. The number of members in warfarin study is also labeled. For instance, NATO consortium has 13 members; ten members from U.S. and three from UK. The experiments illustrate the upper bound results wherein each member defines a different share and acquisition policy for other members (\ie asymmetric relations). In this, each member sends acquisition policy request to consortium members. After a member gets the acquisition request, it reconciles with its share policy and output of negotiation message is returned. The number of messages associated with varying number of selections and conditionals dictated by the members does not require any additional messages. For instance, the acquisition request of a member includes arguments when conditionals are defined (\eg reference data and a threshold value for data-dependent conditionals such as pairwise Jaccard distance), and the result is returned with the negotiation output message. However, the use of the selections and data-dependent conditionals brings additional processing cost as detailed next.

Figure~\ref{fig:members-vs-time} shows the costs associated with the use of CPL selection and data-dependent conditionals. All the members dictate data-dependent conditionals and selections on a single input. The members input size for the data-dependent conditional computations is set to 200 real values. This is the average number of inputs found in members' dataset. Since selections and conditionals reconcile contradictions between acquisition and share policies, they do not require any additional computation overhead and yield a processing time of milliseconds. However, the time associated with varying data-dependent conditionals depend on the protocol of associated secure pairwise algorithm. In our experiments, cosine similarity and intersection size exhibited shorter computation time than Pearson correlation and Jaccard distance. Overall, we found that 25 members compute the metrics less than 18 seconds. Note that the results serve as an upper bound that all members define a set of selections and a data-dependent conditional on one input.

\vspace{-5pt}\subsubsection{Dose Model Benchmarks} Our second series of experiments characterize the impact of CPL on the average time of computing privacy-preserving dose model with varying number of members and dataset sizes. Though the warfarin study includes eight inputs, evaluations are repeated with the input size of 8, 16, 24, 32, and 40 through various dataset sample sizes for completeness. The input and sample size together represents the total dataset shared for a member as a result of the policy agreements. Our experiments show that 80\% of computation overhead is attributed to HE key generation. The cost of the differential privacy takes microseconds, as the members can calculate the (optional) differential private algorithm model at the end of the secure dose protocol. Computations are instrumented to classify the overheads incurred by key generation, encryption, decryption, and evaluation. We next present the costs with and without key generation to study the impact of the number of members and data size.

Figure~\ref{fig:sec-comp-layer} (a-b) presents the computation cost with varying number of members. Each member's dataset includes 5000 data samples which acquired as a result of the policy negotiations. Figure~\ref{fig:sec-comp-layer} (a) presents the cost of the total computation time excluding HE key generation. There is a linear increase in time with the growing number of members. This is the fundamental cost of encryption and evaluation operations dominated by matrix encryption and addition. To profile the generation of key cost, in Figure~\ref{fig:sec-comp-layer} (b), we conducted similar experiments. Each input size cost increases because of the key generation overhead. The increase is quadratic as a number of slots (plaintext elements) are set to square of input size not to lose any data during input conversion. It is important to note that the cost is independent of the member size because a member generates the key only once in a computation of a consortium. We note that the time overhead of key generation is not a limiting factor as members may generate keys before a consortium is established.  

\begin{table*}[t!]
\centering
\renewcommand{\arraystretch}{1.2}
\resizebox{\textwidth}{!}{%
{\small{
\begin{tabular}{|l|l|}
\hline
 \textbf{Member} & \multicolumn{1}{c|}{\textbf{Agreement of policy negotiations}}\\ \hline\hline
U.S. &$\big[$(Race=``Asian'')$\vee$(\textsf{EVALUATE}(age))$\vee$(height \textless 160) $\vee$(weight \textless 65)$\vee$(CYP2C9  \textsf{IN} ( 2*/*2, 2*/*3)$\vee$(Amiodarone=``Y'')$\vee$(Enzyme=``Y'')$\big]$\\\hline
Brasil & $\big[$(Race=``Asian'')$\vee$(height \textless 165)$\vee$(CYP2C9 IN (2*/*2, 2*/*3)$\vee$EVALUATE (Amiodarone)$\vee$(Enzyme=``Y'')$\big]$\\\hline
UK & $\big[$(Race$\neq$``White'')$\vee$(age \textsf{BETWEEN} 20-29 \textsf{AND} \textgreater 80)$\vee$(height\textless 165)$\vee$(60\textless weight \textless 100)$\vee$\textsf{EVALUATE}(CYP2C9)$\vee$(Amiodarone=``Y''), (Enzyme=``Y'')$\big]$ \\\hline
Israel & $\big[$(Race$\neq$ ``White'')$\vee$(height \textless 160cm)$\vee$(weight \textless 60)$\vee$(CYP2C9=3*/*3)$\vee$(Amiodarone=``Y'')$\vee$(Enzyme Inducer =``Y'')$\big]$\\\hline
Taiwan & $\big[$(Race=All)$\vee$(age \textsf{BETWEEN} 20-29)$\vee$(height \textgreater 170)$\vee$(weight \textgreater 65)$\vee$(CYP2C9  \textsf{IN }(1*/*2, 2*/*2, 2*/*3, 3*/*3)$\vee$(VK0RC1=``G/G'')$\vee$(Amiodarone=``Y'')$\vee$(Enzyme=``Y'')$\big]$\\\hline
S. Korea &$\big[$(Race=All)$\vee$ (age \textsf{BETWEEN} 20-29)$\vee$(height \textgreater 165)$\vee$(weight \textgreater 60)$\vee$(CYP2C9  \textsf{IN} (1*/*2, 2*/*2, 2*/*3, 3*/*3)$\vee$(VK0RC1=``G/G'')$\vee$(Amiodarone=``Y'')$\vee$(Enzyme=``Y'')$\big]$\\ \hline
\end{tabular}%
}}}
\caption{An exploration of CPL policies in the global consortium (illustrated as a plain language): Each member defines asymmetric local policy based on its data diversity. The agreement of share and acquisition policies are depicted as a policy clause in a single row. The agreement result of each member for other members is not presented for brevity.}
\label{CPL-policies}
\end{table*}

In Figure~\ref{fig:sec-comp-layer} (c-d), we show the costs associated with different data samples. The number of members in a consortium is set to 20. Similar to the previous experiments, the key generation dominates the computation costs. Our experiments also reported no relationship between the cost and number of samples. That is, even though the size of the data samples increases, the overhead is amortized over the operations on the local statistics of the computations (which is the square matrix of the input size in the warfarin dataset); thus the time of computing dose algorithm converges to the number of dataset inputs. This explains the similar trends observed in plots. 


\subsection{Effectiveness of Policies}
\label{sec:policy-results}
We validate the performance of privacy-preserving dose model quantitatively and qualitatively. For the warfarin study, these are translated to the following questions: How do policies impact the accuracy of members' warfarin dose prediction? (Section~\ref{sec:accuracy}), and Does policies help to prevent the adverse impacts of dose errors on patient health? (Section~\ref{sec:health}).

\vspace{-5pt}

\subsubsection{Implications of CPL on Model Accuracy} 
\label{sec:accuracy}
In our first set of experiments, we validate how well a member prescribe warfarin dose for its local patients and patient's of the consortium members without using CPL. These results are used as a baseline for comparison of varying consortia and data exchange policies throughout. Figure~\ref{fig:policy-results} (a) sought to identify the local algorithm errors (\textbf{P.1}). The errors significantly differ between countries and for the members of the same country (depicted as M$_1$ and M$_2$ in the U.S.). The low results are due to having homogeneous data; all the inputs in these countries have similar traits. For instance, similar age and ethnicity found in a dataset produce over-fitted computation results for its local patients. These findings are validated with use of local algorithms for treatment of other countries' patients. As illustrated in Figure~\ref{fig:policy-results} (b), the dose errors yield significantly high for particular countries' patients. The results indicate that improvements in dose predictions of local patients and members' patients lay in the creation of data exchange policies to increase the patient diversity.

The next experiments measure the impact of CPL in nation-wide (\textbf{P.2}), regional (\textbf{P.3}), NATO-EU (\textbf{P.4}) and global (\textbf{P.5}) consortia. Each member creates a local acquisition policy to acquire the complete data of consortia members (\ie the acquisition policy of a consortium member complies with the share policy of the requested member). We make three major observations. First, varying partnerships yield different dose accuracy. For instance, members of nation-wide consortium get better dose accuracy than their local results. This result is validated through nationwide consortia and a single member (M$_1$) in United States (see Figure~\ref{fig:policy-results} (c)). Second, supporting previous findings, all regional (excluding Asia) and NATO-EU policies decrease the error for both treatment of their patients and the other countries' patients (see Figure~\ref{fig:policy-results} (d-e)). However, Asia consortium results in unexpected dose errors for the treatment of other regions' patients. This is because nation-wide, regional, and NATO-EU policies include patient population having different characteristics; thus the data obtained through policy negotiations better generalize to the dosages. In contrast, Asia collaboration lacks large enough White and Black groups. Third, the global consortium results in higher dose errors when evaluated for particular countries such as Brazil and Taiwan (see Figure~\ref{fig:policy-results} (f)). To conclude, while CPL is effective in reducing dose error of a member, the results highlight the need for the systematic use of CPL through selections and conditionals to obtain better results.

\begin{figure}[t!]
\centering
\includegraphics[width=0.85\columnwidth]{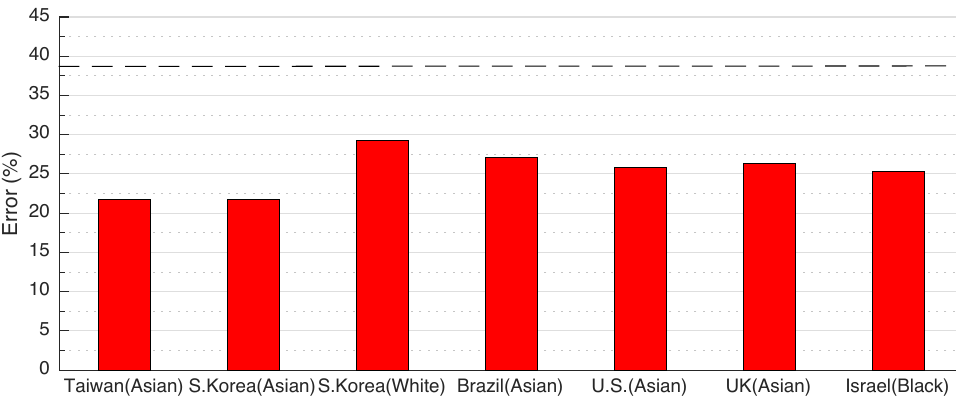}
\caption{Dose accuracy of members using CPL policies defined in Table~\ref{CPL-policies}. Members construct a model per race after they reconcile the policies. The dashed line is the average error found without the use of conditionals and selections in policies.}
\label{fig:filter-similarity}
\end{figure}

In these experiments, each member dictates a different acquisition policy based on its racial groups. Members aim at having an ideal patient population uniformity. To do so, each member defines a local acquisition policy and negotiates it with other members. Each member sets its share policy to conditionals of being in the same consortium and data size greater than 200; thus, the policy of each member is asymmetric. Table~\ref{CPL-policies} shows the simplified notation of the policy agreements in the global consortium. For instance, a member having a small number of white patients defines selections to solely acquire that group and a member having large enough patients for all genotypes sets data-dependent conditionals to obtain patient inputs that are not similar in its data samples (\eg acquires different genotypes). Figure~\ref{fig:filter-similarity} presents a subset of results on dose errors per patient race. The errors of the other races yield similar for each member. The results without CPL conditionals and selections are plotted as a dashed line for comparison. We find that members can improve the dose accuracy with the use of policies. We note that the use of different data-dependent conditionals defined in \textsf{evaluate} does not result in statistically significant accuracy gain.

\vspace{-5pt}\subsubsection{Implications of CPL on Patient Health}
\label{sec:health}
We examine the impact of the dose errors found in the previous section to better quantify the effectiveness of policies on patient health. 

To identify the adverse effects of warfarin, we use a clinical study to evaluate the clinical relevance of prediction errors~\cite{patien-driven} and a medical guide to identify the consequences of over- and under-prescriptions~\cite{fdaWarfarinRisk}. We define errors that are inside and outside of the warfarin safety window, and the under- or over prescriptions. We consider weekly errors for each patient because using weekly values eliminates the errors posed by the initial (daily) dose. The weekly dose is in the safety window if an estimated dose falls within 20\% of its corresponding clinically-deduced value~\cite{international2009estimation, kimmel2013pharmacogenetic}. The deviations falling outside of the safety window is an under- or over prescriptions, and cause health-related risks.

Table~\ref{table:patient-health} presents the percentage of patients falls in safety window, over- and -under prescriptions with varying policies of a member. We find that use of CPL increases the number of patients in the safety window. For instance, a member has 43.4\% patient with using its local data (single source model), and the member increases the percentage of patients in a safety window with varying consortia and policies, for instance, it is 52.4\% in the nation-wide consortium. We conclude that CPL might be useful in preventing errors that introduce health-related risks.

\begin{table}[t!]
\centering
\setlength{\tabcolsep}{4pt}
\renewcommand{\arraystretch}{1}
\resizebox{\columnwidth}{!}{%
{\small{
\begin{tabular}{|l|c|c|c|c|c|}
\hline
\textbf{Consortium}& \textbf{U} & \textbf{SW} & \textbf{O} & \textbf{Selections} & \textbf{Conditionals} \\ \hline\hline
Single Source & 37.7\% & 43.4\% & 18.8\% & \xmark & \xmark \\\hline
Nation-wide & 18.9\% & 52.3\% & 28.8\% & \cmark & \cmark \\\hline
NATO  & 19.3\% & 51.5\% & 29.2\% & \cmark & \cmark \\\hline
Regional  & 19\% & 51.3\% & 29.7\% & \cmark & \cmark \\\hline
Global  & 21.2\% & 46.8\% & 32\% & \cmark & \cmark \\ \hline
\end{tabular}%
}}}
\caption{Impact of policies on health-related risks: Results are from a global consortium patients using policy agreement of a member located in the U.S. The member uses the policy defined in Table~\ref{CPL-policies}. (U: Under-prescription, SW: Safety Window, O: Over-prescription)}
\label{table:patient-health}
\end{table}

\section{Limitations and Discussion}
One requirement for correctly interpreting the CPL policies is a shared schema for solving the compatibility issues among members. For instance, members may interpret the data columns (\eg column names and types) differently or may not have the information about consortium members (\eg membership status of an alliance). CPL implements a shared schema describing column names, their types, and explanations of data fields as well as consortium-specific information. Members can negotiate the schema similar to the policy negotiations and revise the schema based on the schema of a negotiation initiator.

CPL provides a set of data-dependent statistical functions (\eg cosine similarity) to compute pairwise statistics among member's local data. However, there might be a need for other functions that help members decide their data exchange policies. For example, data exchange among finance companies may require calculating the similarity between data distributions. Future work will investigate the integration of different data-dependent statistics into CPL.

Lastly, we did not focus much on the reasons of policy impacts on the prediction success of the dose algorithm and its adverse outcomes on patient health over time. While our evaluation results showed that members could express both complex relations and constraints on the data exchange through CPL policies, members require establishing true partnerships to improve the prediction model accuracy. While this explanation matches both our intuition and the experimental results, a further domain-specific formal analysis is needed. We plan to pursue this in future work.

\section{Related Work}
\label{sec:related-work}
Policy has been used in several contexts as a vehicle for representing configuration of secure groups~\cite{mcdaniel2006methods}, network management~\cite{riekstin2016orchestration}, threat mitigation~\cite{freudiger2015controlled}, access control~\cite{duan2016automated}, and data retrieval systems~\cite{elnikety16thoth}. These approaches define a schema for their target problem and do not consider the challenges in secure data exchange. In contrast, \Curie defines a formal policy language to dictate the data exchange requirements of members and enforces the agreement in collaborative ML settings.

On the other hand, secure computation on sensitive proprietary data has recently attracted attention. Federated learning~\cite{geyer2017differentially,smith2017federated}, anonymization~\cite{el2013secure}, multi-site statistical models~\cite{dankar2015privacy}, secure multi-party computation~\cite{lindell2009secure}, and secure and differentially-private multi-party computation~\cite{abbas2017PAC} have started to shed light on this issue. Such techniques have been used both for training and classification phases in deep learning~\cite{shokri2015privacy}, clustering~\cite{graepel2012ml}, and decision trees~\cite{bost2015machine}. To allow programmers to develop such applications,  secure computation programming frameworks and languages are designed for general purposes~\cite{henecka2010tasty, rastogi2014wysteria, ohrimenko2016oblivious, bogdanov2016rmind, el2013secure}. However, these approaches do not consider complex relationships among members and assume members share their all data or nothing. We view our efforts in this paper to be complementary to much of these works. CPL can be integrated into these frameworks to establish partnerships and manage data exchange policies before a computation starts.

\section{Conclusions}
We presented \Curie which provides a novel policy language called CPL to define the specifications of data exchange requirements securely for use in collaborative learning settings. Members can assert who and what to exchange separately for data sharing and data acquisition policies. This allows members to efficiently dictate their policies in complex and asymmetric relationships through selections, conditionals, and pairwise data-dependent statistics. We validated \Curie in an example real-world healthcare application through varying policies of consortia members. A secure multi-party and (optional) differentially-private model is implemented to illustrate the policy/performance trade-offs. \Curie allowed 50 different members to efficiently compute a privacy-preserving model using 5K data samples with 40 inputs in less than a minute. We also showed how an algorithm with effective use of data exchange policies could improve the accuracy of the dose prediction model.

Future work will investigate the use of \Curie in other 
collaborative learning settings exploring different statistics for data-dependent conditionals and explore its performance trade-offs by integrating it into other off-the-shelf secure computation frameworks.

\section*{Acknowledgment}
Research was sponsored by the Army Research Laboratory and was accomplished under Cooperative Agreement Number W911NF-13-2-0045 (ARL Cyber Security CRA). This work is also partially supported by US National Science Foundation (NSF) under the grant numbers NSF-CNS-1718116 and NSF-CAREER-CNS-1453647. The statements made herein are solely the responsibility of the authors. The views and conclusions contained in this document are those of the authors and should not be interpreted as representing the official policies, either expressed or implied, of the Army Research Laboratory or the U.S. Government. The U.S. Government is authorized to reproduce and distribute reprints for Government purposes notwithstanding any copyright notation here on.

\bibliographystyle{ACM-Reference-Format}
\bibliography{acsac-18/refs/abbas,acsac-18/refs/secure-data-share}

\appendix

\setcounter{figure}{0}
\setcounter{equation}{0}
\setcounter{table}{0}

\section{Curie Policy Language} \label{sec:CPL}
This section presents the Backus Naur Form of Curie data exchange policy language.
\setlength{\grammarparsep}{4pt plus 1pt minus 1pt} 
\setlength{\grammarindent}{8em} 

{\footnotesize{
\begin{grammar}
<curie_policy> ::= <statements> 

<statements> ::= <statement> `;' [<statements>]

<statement> ::= <share_clause> 
\alt  <acquire_clause>
\alt  <attribute>  
\alt  <sub_clause>
\end{grammar}}}

\noindent{; share clauses defined as follows:}

{\footnotesize{
\begin{grammar}
<share_clause> ::= `share' `:' [<members>] `:'  [<conditionals>] \\`::' <selections>  
\end{grammar}}}

\noindent{; acquisition clauses defined as follows:}

{\footnotesize{
\begin{grammar}
<acquire_clause> ::= `acquire' `:' [<members>] `:'  [<conditionals>] `::' <selections>  
\end{grammar}}}

\noindent{; attributes are defined as follows:}

{\footnotesize{
\begin{grammar}
<attribute> ::= <identifier> `:=' `<' <value> `>'  
\alt  <identifier> `:=' `<' <value_list> `>'  
\end{grammar}}}

\noindent{; user defined sub-clauses defined as follows:}
{\footnotesize{
\begin{grammar}
<sub_clause> ::= <tag> `:'  [<conditionals>] `::' <selections>  
\end{grammar}}}

\noindent{; conditionals including data-dependent functions defined as follows:}

{\footnotesize{
\begin{grammar}
<conditionals> ::= <var>`='<value> [`,' <conditionals>]
\alt `evaluate' `(' <data_ref> `,' <alg_arg> `,' \\<threshold_arg> `)' [`,' <conditionals>] | `'

<selections> ::= <filters> 
\alt  <tag> 

<filters> ::= <filter> [`,' <filters>] 

<filter> ::= <var> <operation> <value> | `'

<data_ref> ::=  `\&' <identifier> 

 <alg_arg> ::=  <algorithms> 

<algorithms> ::=  `Intersection size' \alt `Jaccard index' \alt `Pearson correlation' \alt `Cosine similarity'

<threshold_arg> ::=  <floating_point_number> 

<operation> ::= `=' | `<' | `>' | `!=' | in | 

<value_list> ::= `\{' <value> `\}' [`,' <value_list>] 

<members> ::= <member> [`,' <members>] 

<member> ::= <identifier> | `'

\end{grammar}}}

\noindent{; for completeness, trivial items defined as follows:}

{\footnotesize{
\begin{grammar}

<identifier> ::= <word>

<var> ::=  `\$' <identifier> 

<value> ::=  <string> 

<tag> ::= <word>  

<string> ::= `"' <stringchars> `"'

<stringchars> ::= <stringletter> [ <stringchars>]

<stringletter> ::= 0x10 | 0x13|0x20| ... | 0x7F 

<word>      ::=  <char> [ <word> ]

<char>        ::= <letter>  |  <digit>

<letter> ::= 'A' | 'B' | ... | 'Z' | 'a' | 'b' | ... | 'z' | 0x80 | 0x81 | ... | 0xFF

<digit> ::= '0' | '1' | ... | '9'

<floating_point_number> ::=  <decimal_number> '.' [<decimal_number>]

<decimal_number> ::= <digit> [ <decimal_number> ]


\end{grammar}}}

\section{Differentially-Private Dose Algorithm}
\label{sec:dif-privacy}
We presented how members compute a privacy-preserving dose model on negotiated data through their policies. In this section, we consider individual privacy that allows a member to guarantee no information leakage on the targeted individual (\ie patient) involved in the computation. Specifically, while members compute a secure dose model using the data obtained as a result of their policy negotiations, they also ensure that an adversary cannot infer whether any particular individual is included in computations to build the dose algorithm. In warfarin study, this corresponds to a differentially-private secure dose algorithm on shared data. 

To implement a differentially-private secure algorithm, we use a functional mechanism technique~\cite{zhang2012functional,wu2015revisiting}. The technique accepts a dataset ($\mathcal{D}$), an objective function ($f_\mathcal{D}(\eta)$), and  a privacy budget ($\epsilon$) as an input and returns $\epsilon$-differentially-private coefficients $\widetilde{\eta}$ of an algorithm. The intuition behind the functional mechanism is perturbing the objective function of the optimization problem. Perturbation includes both sensitivity analysis and Laplacian noise insertion as opposed to perturbing the results via differentially-private synthetic data generation. 

To inter-operate the functional mechanism with the secure dose protocol, members convert each column from [min, max] to [-1,1] before negotiation starts. This processing ensures that sufficient noise is added to the objective function on negotiated data. Then, members proceed with the protocol. At the final stage of the secure algorithm protocol, a member gets clear statistics of $\mathcal{O}= \mathcal{X}^\intercal \mathcal{X}$ and $\mathcal{V}= \mathcal{X}^\intercal \mathcal{Y}$ and input dimension $d$ that is the size of $\mathcal{O}$ or $\mathcal{V}$. These statistics are the exact quantities that are minimized in the objective of the functional mechanism~\cite{zhang2012functional}. Using these statistics, a member may (optionally) compute $\epsilon$-differential private secure algorithm without any additional data exchange and computational overhead.

\begin{figure}[t!]
\begin{center}
\includegraphics[width=0.75\columnwidth]{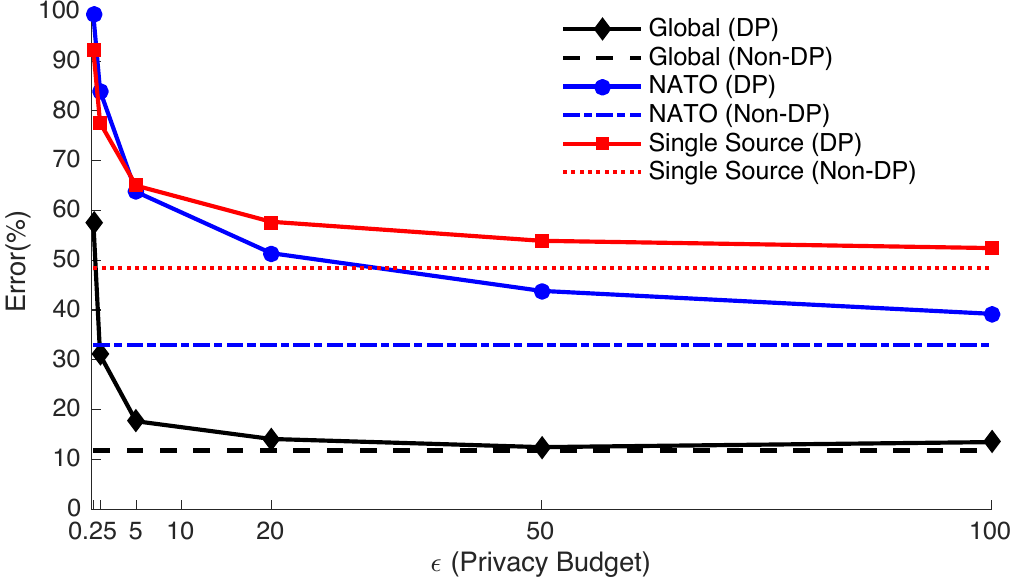}
\end{center}
\caption{Non-private secure algorithm (Non-DP) vs. differentially-private secure algorithm (DP) performance of a member in U.S. measured against various policies depicted in Figure~\ref{fig:policy-results}.}
\label{fig:dif-privacy}
\end{figure}

\vspace{2pt}\noindent\textbf{Differential Privacy Results.} To protect individual privacy in secure dose algorithm, members may compute the differentially-private secure algorithm on their negotiated data. This section presents the results of using the differential-private secure algorithm (DP) instead of using secure dose algorithm (Non-DP). To establish a baseline performance, we constructed non-private secure algorithms of a member. We then build the differential-private secure algorithm for different privacy budgets ($\epsilon$ =  0.25, 1, 5, 20, 50 and 100). Finally, we compare the results of two algorithms through different policies of a member. Figure~\ref{fig:dif-privacy} shows the results of a member in the U.S. that applies both algorithms to predict the dosage. The algorithms are constructed for the single source, NATO, and global consortia. In this, the member dictates acquisition policy for complete data and other members complies with their share policy. The average error over 100 distinct model for each budget value is reported. The use of DP degrades the accuracy as the $\epsilon$ value increases. For instance, the accuracy improvement obtained through NATO policy over single source degrades with the privacy budget less than or equal to 20. We note that other consortia and policies with use of selections and conditionals show similar effect on the dose accuracy. 

\section{Analysis of the Dose Algorithm}
\label{sec:securityAnalysis}
We present security and privacy guarantees of the dose algorithm provided to all members through the share of encrypted integrated statistics, ($\mathcal{O}_{i}= \mathcal{X}^\intercal \mathcal{X}$ and $\mathcal{V}_{i}= \mathcal{X}^\intercal \mathcal{Y}$ matrices). Since all data exchange among parties is encrypted through the use of HE, the security of the algorithm against any adversary outside the authorized parties is based on the underlying HE cryptosystem. 

\vspace{2pt}\noindent\textbf{An adversary not involving session initiator.} Assume for now that a session initiator does not collude with other parties. 
Loosely speaking, since all computations are performed on the encrypted data, none of the parties learn anything about other parties' input. 

We consider a party $P_{i+1}$ in Figure~\ref{fig:secure-dose-algorithm}. The party $P_{i+1}$ has the public key generated by the session initiator $K_i$, the encryption of local statistics of previous parties $M_i=(E(\mathcal{O}_i)_{K},E(\mathcal{V}_i)_{K})$. Its input is $(\mathcal{V}_{i+1},\mathcal{O}_{i+1})$ and its output is $M_{i+1}=(E(\mathcal{O}_i+\mathcal{O}_{i+1}),E(\mathcal{V}_i+\mathcal{V}_{i+1}))$. A simulator $S$ selects random values for its own inputs $(\mathcal{V}'_{i+1},\mathcal{O}'_{i+1})$ and encrypts them using the public key published by the session initiator. Then, the simulator $S$ performs the homomorphic operation on the received message $M_i$ and outputs $M'_{i+1}=(E(\mathcal{O}_i+\mathcal{O}'_{i+1})_{K},E(\mathcal{V}_i+\mathcal{V}'_{i+1})_{K})$. Here, we assume the underlying HE is semantically secure. Therefore, the output of the simulator $M'_{i+1}$ is computationally indistinguishable from output of the real execution of the protocol $M_{i+1}$ for every input pairs. Therefore, using the definition in~\cite{goldreich2009foundations} the protocol privately computes the function in the presence of one semi-honest corrupted party. The extension to multi-corrupted semi-honest adversaries is straightforward as the only difference is the view of a subset of parties having many encrypted messages. Since the semantic security of the underlying HE is hold for any pair of these many encrypted messages, no information leaks about the corresponding plaintexts. 

\vspace{2pt}\noindent\textbf{Adversary involving session initiator.} We consider the case when the session initiator is corrupted. The corrupted parties including session initiator can infer the input of an honest party if the predecessor (previous party) and successor (next party) of an honest party are both corrupted. We consider the possible cases for data leakage: (1)
\emph{2-party:} The session initiator is corrupted, and another party is honest. In this case, predecessor and successor of the honest party are both the corrupted session initiator. Therefore, the input of honest party is learned by the corrupted party, (2) \emph{3-party}: A corrupted session initiator is either predecessor or successor; thus it can learn inputs of the one of the honest party only if another party is corrupted, and (3) \emph{n-party ($n>3$)}: To learn an honest party's input, at least two parties must be corrupted and placed in previous and next of the honest party. 

While the individual raw data of members does not leak, the risk of inappropriate disclosures from local summary statistics exists in some extreme cases~\cite{el2013secure}. Consider the exchange of plain matrix $V_{i}= {X}^\intercal Y$ among two parties; a party may use the extreme values found in $V_{i}$ to identify particular patients. For instance, in dose algorithm, taking inducers such as Rifadin and Dilantin could indicate high dose prescriptions. If the values of $V_{i}$ are high, then a party may infer a patient that takes enzyme inducers and the presence of high dosage warfarin intake. Similarly, exchange of $O_{i}= {X}^\intercal X$ may leak information about the number of observations and represent the number of 0s or 1s in a column. For instance, for the former first entry in the matrix, ${X}^\intercal X$,  gives the total number of patients. For the latter, $(X^\intercal X)_{j, j}$ gives the number of 1s in the column. This type information lets a party infer knowledge, particularly when binary inputs (\eg use of the medicine) are used.

\section{Curie Deployment Details}
\label{sec:appendix-members}
We use a dataset collected by the International Warfarin Pharmacogenetics Consortium (IWPC), to date the most comprehensive database containing patient data collected from 24 medical institutions from 9 countries~\cite{international2009estimation}. 
The dataset does not include the name of the medical institutions, yet there is a separate ethnicity dataset provided for identifying the genomic impacts of the algorithm. We use the race (reported by patients) and race categories (defined by the Office of Management and Budget) to predict the country of a patient\footnote{The authors indicated via personal communication that they cannot provide the exact name of the institutions due to the privacy concerns.}. For instance, we consider a medical institution with a high number of Japanese race is located in Japan. We use subsets of patient records that have no missing inputs for accurate evaluation. We split the dataset into two cohorts: training cohort is used to learn the algorithm, and validation cohort is used to assign dose to the new patients based on the consortia and data exchange policies.

\end{document}